\documentclass[a4paper,11pt]{article}
\usepackage[dvips]{graphicx}
\usepackage{amssymb}
\usepackage{amsmath,amscd}
\usepackage{cite}

\begin{document}
 
\input arrow

\title{Hidden Supersymmetry in Dirac Fermion Quasinormal Modes of Black Holes}
\author{V. K. Oikonomou\thanks{
Vasilis.Oikonomou@mis.mpg.de}\\
Max Planck Institute for Mathematics in the Sciences\\
Inselstrasse 22, 04103 Leipzig, Germany} \maketitle

\begin{abstract}
We connect the quasinormal modes corresponding to Dirac fermions in various curved spacetime backgrounds to an $N=2$ supersymmetric quantum mechanics algebra, which can be constructed from the radial part of the fermionic solutions of the Dirac equation. In the massless fermion case, the quasinormal modes are in bijective correspondence with the zero modes of the fermionic system and this results to unbroken supersymmetry. The massive case is more complicated, but as we demonstrate, supersymmetry remains unbroken even in this case.  

\end{abstract}

\section*{Introduction}

Black holes in equilibrium are generally speaking, simple objects
due to the very few parameters that are needed to describe them \cite{kokkotas,review,review1,roman}.
However, it is physically impossible to have an isolated black hole
in nature, due to the fact that matter always exists around them,
interacting directly with the black hole. This leads us to the
conclusion that a black hole is always in a some sort of perturbed
state, with more parameters needed to describe it, than in the
unperturbed state. Quasinormal modes \cite{kokkotas,review,review1,roman,ker,ker1,ker2,ker3,ker4,ker5,kern1,kern2,kern3,rn1,rn2,rn3,rn4,rn5,rn6,rn7,schwarzchild,schwarzchild1,schwarzchild2,schwarzchild3,quasidirac,quasidirac1,quasidirac2,quasidirac3,quasidirac4,quasidirac5,quasidirac6,quasidirac7,kanti} describe a long
period of damped proper oscillations of gravitational waves, and
are extremely important in various physical phenomena. The direct
observation of black holes is actually based on quasinormal modes,
with the most dominating one being the fundamental mode, namely
the lowest frequency in the spectrum. The quasinormal modes are the characteristic sound of black holes and being such, matter field perturbations of such extreme gravitational backgrounds can reveal many important properties of black holes. 

\noindent The perturbation of a black hole can be achieved either by
directly perturbing the gravitational background or by simply
adding matter or gauge fields in the black hole spacetime \cite{roman}. In this paper we shall
shall use the latter approach, in the linear approximation, which suggests
that the field has no back-reaction on the metric. Particularly, we shall study Dirac fermion systems around various black hole and spacetime environments and study when the system possesses an unbroken hidden supersymmetry. Supersymmetry has been connected to quasinormal modes spectra in the past, but in a different context \cite{quasinormalsupersymmetry1,quasinormalsupersymmetry2,quasinormalsupersymmetry3,quasinormalsupersymmetry5,quasinormalsupersymmetry6}. Most of these works studied bosonic quasinormal modes and their relation to supersymmetry. We study the zero modes of the fermionic system and directly relate these to the quasinormal modes. As we shall see, the zero modes and quasinormal modes have a bijective correspondence, a fact that can actually be very crucial for supersymmetry to be unbroken. The specific type of supersymmetry that we found is an $N=2$ supersymmetric quantum mechanics \cite{witten,thaller,susyqm,susyqm1,susyqmarxiv,susyqmarxiv1,susyqmarxiv2,susyqmarxiv3,susyqmarxiv4,pluskai1,pluskai2,pluskai3} (shortened to SUSY QM hereafter) with zero central charge. The aforementioned supersymmetry is inherent to many gravitational systems \cite{oikonomou1,oikonomou2}. In this paper the focus is on supersymmetries in gravitational backgrounds. We exploit the existence of the quasinormal modes in this backgrounds in order to establish the fact that the Witten index of the corresponding underlying supersymmetric algebra is zero, with the kernels of the corresponding operators being non-empty and consequently, supersymmetry is unbroken.

\noindent This paper is organized as follows. In section 1 we study the supersymmetric underlying structure for the case of a Kerr black hole, with the latter being the most physically interesting black hole satisfying the Einstein's equations. In section 2, we study the massless and massive Dirac fermion in Kerr-Newman, Reissner-N\"{o}rdstrom, Schwarzschild gravitational backgrounds. The massive case proves to be much more complicated compared to the massless case, but still, supersymmetry remains unbroken. In section 3, the focus is on three different spacetimes, namely, the D-dimensional de Sitter, Kerr-Newman-de Sitter and Reissner-Nordstr\"{o}m-anti-de Sitter spacetimes, finding the same results as in the previous sections. In reference to the Reissner-Nordstr\"{o}m-anti-de Sitter spacetime, we find that this fermionic system has two $N=2$ SUSY QM algebras. In section 4, we present some physical and mathematical implications of the SUSY QM on the fermionic system and also to the fibre bundle structure of the spacetime. In addition, we study the impact of compact radial perturbations to the Witten index of the SUSY QM algebra, in the case the spacetime is maximally symmetric. Moreover, we address the question if there can be a higher extended supersymmetry underlying these $N=2$ supersymmetries we found. The conclusions follow in section 5.

\section{Massless Fermion Quasinormal modes for a Kerr Black Hole}

We shall study first the quasinormal modes of a Dirac fermion in a Kerr black hole background. Particularly, we focus on the massless fermion case. As we mentioned earlier, the Kerr black hole is one of the most important four dimensional black hole solutions, since realistic astrophysical black holes are rotating with negligible electric charge and in addition, the quasinormal modes stemming from such a background are very important for observations of gravitational waves \cite{gravitationalwaves,gravitationalwaves1,detectgravitywaves,detectgravitywaves1,detectgravitywaves2,detectgravitywaves3} (and more likely gravity waves will come from such objects).

\noindent Following references \cite{penrose,quasidirac,quasidirac1,quasidirac2,quasidirac3,quasidirac4,quasidirac5,quasidirac6,quasidirac7} and employing the Newman-Penrose formalism, the massless Dirac equation in the null tetrad basis reads:
\begin{equation}\label{spinor}
i\gamma^{\mu}\nabla_{\mu}\Psi=0,
\end{equation}
with the covariant derivative being equal to \cite{Jost,Jost1,Nakahara,eguchi},
\begin{equation}\label{covder}
\nabla_{\mu}=\partial_{\mu}-\frac{i}{4}\omega_{\mu{\,}{\,}{\,}b}^a\eta_{ca}\gamma^{c{\,}b}.
\end{equation}
The spin connection $\omega_{\mu{\,}{\,}{\,}b}^a$ on the pseudo-Riemannian manifold, satisfies the following equation:
\begin{equation}\label{vielbeinan1}
\partial_{\mu}e^{a}_{\nu}+\omega_{\mu{\,}{\,}{\,}b}^ae^{b}_{\nu}-\Gamma_{\mu{\,}\nu}^{\sigma}e_{\sigma}^{a}=0.
\end{equation}
The four dimensional Kerr background spacetime has the following metric:
\begin{align}\label{metrickerrblackhole}
&\mathrm{d}s^2=-\Big{(}1-\frac{2Mr}{\rho^2}\Big{)}\mathrm{d}t^2-\Big{(}\frac{4Mra\sin^2\theta}{\rho^2}\Big{)}\mathrm{d}t\mathrm{d}\theta \\ \notag &+\frac{\rho^2}{\Delta}
\mathrm{d}r^2
+\rho^2\mathrm{d}\theta^2+\Big{(}r^2+a^2+\frac{2Mra\sin^2\theta}{\rho^2}\Big{)}\sin^2\theta \mathrm{d}\phi^2,
\end{align}
with, the parameters $\rho^2$, $\Delta$, $a$ appearing above, being equal to:
\begin{align}\label{parrev2}
& \rho^2=r^2+a^2\cos^2\theta, \\ \notag &
\Delta=r^2+a^2-2Mr, \\ \notag &
a=\frac{J}{M}.
\end{align}
For later purposes, we introduce the parameters $\bar{\rho},\bar{\rho}^*$ to be:
\begin{equation}\label{paramdef}
\bar{\rho}=r+ia\cos\theta,{\,}{\,}{\,}\bar{\rho}^*=r-ia\cos\theta,
\end{equation}
such that $\rho^2=\bar{\rho}\bar{\rho}^*$. In addition, we define the following operators, which shall frequently be used in this section:
\begin{align}\label{basoper1}
&\mathcal{D}_0=\partial_r+i\frac{K}{\Delta},
{\,}{\,}{\,}\mathcal{D}_0^{\dag}=\partial_r-i\frac{K}{\Delta}\\
\notag & \mathcal{L}_0=\partial_{\theta}+Q,
{\,}{\,}{\,}\mathcal{L}_0=\partial_{\theta}-Q,
\end{align}
with $K=(r^2+a^2)\omega+am$ and $Q=a\omega \sin \theta
+\frac{m}{\sin \theta}$, and also,
\begin{align}\label{basoper2}
&\mathcal{D}_n=\partial_r+i\frac{K}{\Delta}+2n\frac{r-M}{\Delta},
{\,}{\,}{\,}\mathcal{D}_n^{\dag}=\partial_r-i\frac{K}{\Delta}+2n\frac{r-M}{\Delta}\\
\notag & \mathcal{L}_n=\partial_{\theta}+Q+\frac{n}{\tan
\theta}-\frac{ina\sin \theta}{\bar{\rho}},
{\,}{\,}{\,}\mathcal{L}_n^{\dag}=\partial_{\theta}-Q+\frac{n}{\tan
\theta}+\frac{ina\sin \theta}{\bar{\rho}^*}.
\end{align}
Then, following \cite{quasidirac,quasidirac1,quasidirac2,quasidirac3,quasidirac4,quasidirac5,quasidirac6,quasidirac7} the Dirac equation (\ref{spinor}) in the Kerr spacetime can be cast in the form:
\begin{align}\label{recdiraceqn}
&\mathcal{D}_0f_1(r,\theta )+\frac{1}{\sqrt{2}}\mathcal{L}_{\frac{1}{2}}f_2(r,\theta )=0\\
\notag &\Delta
\mathcal{D}_{\frac{1}{2}}^{\dag}f_2(r,\theta )-\sqrt{2}\mathcal{L}^{\dag}_{\frac{1}{2}}f_1(r,\theta )=0
\\
\notag &
\mathcal{D}_{0}g_2(r,\theta )-\frac{1}{\sqrt{2}}\mathcal{L}^{\dag}_{\frac{1}{2}}g_1(r,\theta )=0
\\
\notag &\Delta
\mathcal{D}_{\frac{1}{2}}^{\dag}g_1(r,\theta )+\sqrt{2}\mathcal{L}_{\frac{1}{2}}f_1(r,\theta )=0,
\end{align}
where $D_{\frac{1}{2}}$ and $\mathcal{L}_{\frac{1}{2}}$, can be
deduced from Eq.(\ref{basoper2}). In order to make contact with
the quasinormal modes spectrum, we separate the above functions
into radial and angular parts, as follows:
\begin{align}\label{radiaangular}
&f_1(r,\theta)=R_{-\frac{1}{2}}(r)S_{-\frac{1}{2}}(\theta),\\
\notag &f_2(r,\theta)=R_{\frac{1}{2}}(r)S_{\frac{1}{2}}(\theta),\\
\notag
&g_1(r,\theta)=R_{\frac{1}{2}}(r)S_{-\frac{1}{2}}(\theta),\\
\notag &g_2(r,\theta)=R_{-\frac{1}{2}}(r)S_{\frac{1}{2}}(\theta).
\end{align}
By substituting Eq.(\ref{radiaangular}), to
Eq.(\ref{recdiraceqn}), we obtain the angular part of the Dirac equation:
\begin{align}\label{finalan111g}
&\mathcal{L}_{\frac{1}{2}}S_{\frac{1}{2}}(\theta)=-\lambda
S_{-\frac{1}{2}}(\theta)\\ \notag &
\mathcal{L}_{\frac{1}{2}}^{\dag}S_{-\frac{1}{2}}(\theta)=\lambda
S_{\frac{1}{2}}(\theta),
\end{align}
and the radial part of the Dirac equation:
\begin{align}\label{finalrad}
&\mathcal{D}_{0}R'_{-\frac{1}{2}}(r)=\frac{\lambda}{\sqrt{\Delta}}R_{\frac{1}{2}}(r)\\
\notag &
\mathcal{D}_{0}^{\dag}R_{\frac{1}{2}}(r)=
\frac{\lambda}{\sqrt{\Delta}}R'_{-\frac{1}{2}}(r),
\end{align}
with $R'_{-\frac{1}{2}}(r)=\frac{\sqrt{2}}{\sqrt{\Delta}}R_{-\frac{1}{2}}(r)$. Equations (\ref{finalan111g}) and (\ref{finalrad}) correspond to the angular and radial part of the Dirac equation respectively.
Eliminating $S_{\frac{1}{2}}$ or $S_{-\frac{1}{2}}$ from the above equation, and upon defining $u=\cos \theta$, we can find that the angular equation can be written as (with $s=\pm\frac{1}{2}$):
\begin{equation}\label{angularpareal}
\frac{\mathrm{d}}{\mathrm{d}u}\Big{(}(1-u^2)\mathrm{d}S_s(\theta){\mathrm{d}u}\Big{)}+\Big{(}(a\omega u)^2-2{\,}a\omega {\,}s{\,} u+s+A_{lm}-\frac{m+s{\,}u}{1-u^2}\Big{)}S_s=0,
\end{equation}
with $A_{lm}=\lambda^2+2{\,}m{\,}a{\,}a-(a\omega)^2$. The parameter $\lambda$ is real, but our results are indifferent to whether $\lambda$ is real or not. This would change slightly our notation, but the results would be the same.
Equations (\ref{finalan111g}) and (\ref{finalrad}), will be our starting point of our analysis. We shall see that the solutions of these equations are related to a supersymmetric Hilbert space.
The above equation (\ref{finalrad}) (using the tortoise coordinate $x$, defined as $\frac{\mathrm{d}r}{\mathrm{d}x}=\frac{\Delta \omega}{K}$), can be transformed to the following Schroedinger like equation \cite{quasidirac,quasidirac1,quasidirac2,quasidirac3,quasidirac4,quasidirac5,quasidirac6,quasidirac7}:
\begin{equation}\label{schroedingerlike11234}
\frac{\mathrm{d}^2Z_{\pm}}{\mathrm{d}x^2}+\Big{(}\omega^2-V_{\pm}(x)\Big{)}Z_{\pm}=0,
\end{equation}
with $Z_{\pm}=P_{\frac{1}{2}}\pm P_{-\frac{1}{2}}$ and $V_{\pm}(x)=\lambda^2\frac{\Delta}{\bar{K}^2}\pm \lambda\frac{\mathrm{d}}{\mathrm{d}x}\Big{(}\frac{\sqrt{\Delta}}{\bar{K}}\Big{)}$. Equation (\ref{schroedingerlike11234}) is the quasinormal modes master equation for a fermion field in Kerr spacetime. We are not interested in solving the master equation, there are quite rigorous techniques for doing that \cite{kokkotas,review,review1,roman,ker,ker1,ker2,ker3,ker4,ker5,kern1,kern2,kern3,rn1,rn2,rn3,rn4,rn5,rn6,rn7,schwarzchild,schwarzchild1,schwarzchild2,schwarzchild3,quasidirac,quasidirac1,quasidirac2,quasidirac3,quasidirac4,quasidirac5,quasidirac6,quasidirac7}. Our main interest is whether the spectrum in terms of $\omega$ is discrete or continuous. Quasinormal modes are solutions of the above equation, with the wave functions satisfying
certain boundary conditions at the horizon and at infinity. In
addition, the quasinormal modes corresponding to Kerr black holes
form a countable set of discrete frequencies. The boundary conditions are very crucial in order to define a trace class operator, which be valuable to us in the following. 
Based on equations (\ref{finalan111g}) and (\ref{finalrad}), we can construct an $N=2$ supersymmetric quantum algebra acting on the fermionic solutions. We define the matrix $\mathcal{D}_K$,
\begin{equation}\label{susyqmrn567m}
\mathcal{D}_K=\left(%
\begin{array}{cc}
 D_0 & \frac{\lambda}{\sqrt{\Delta}}
 \\ \frac{\lambda}{\sqrt{\Delta}} & D_0^{\dag}\\
\end{array}%
\right),
\end{equation}
acting on the vector:
\begin{equation}\label{ait34e11}
|\phi ^{-}_K \rangle =\left(%
\begin{array}{c}
  R'_{-\frac{1}{2}}(r) \\
  R_{+\frac{1}{2}}(r) \\
\end{array}%
\right).
\end{equation}
Replacing the operators $D_0$ from equation (\ref{basoper2}), $\mathcal{D}_K$ is equal to:
\begin{equation}\label{susyqmrn12m}
\mathcal{D}_K=\left(%
\begin{array}{cc}
 \Big{(}\partial_r+i\frac{K}{\Delta}\Big{)} & \frac{\lambda}{\sqrt{\Delta}}
 \\ \frac{\lambda}{\sqrt{\Delta}} & \Big{(}\partial_r-i\frac{K}{\Delta}\Big{)}\\
\end{array}%
\right).
\end{equation}
We can easily obtain $\mathcal{D}_K^{\dag}$ which is equal to (note that $K$ contains $\omega$ which is complex):
\begin{equation}\label{susyqmrnmd2ag1}
\mathcal{D}_K^{\dag}=\left(%
\begin{array}{cc}
 \Big{(}\partial_r-i\frac{K^*}{\Delta}\Big{)} & \frac{\lambda}{\sqrt{\Delta}}
 \\ \frac{\lambda}{\sqrt{\Delta}} &  \Big{(}\partial_r+i\frac{K^*}{\Delta}\Big{)}\\
\end{array}%
\right),
\end{equation}
acting on,
\begin{equation}\label{aigfgfhte21}
|\phi^{+}_K\rangle =\left(%
\begin{array}{c}
   \Big{(}R'_{-\frac{1}{2}}(r)\Big{)}^* \\
 \Big{(}R_{+\frac{1}{2}}(r)\Big{)}^* \\
\end{array}%
\right).
\end{equation}
Obviously, each quasinormal mode satisfies the zero mode equation of $\mathcal{D}_{K}$. Therefore, we could say that the quasinormal modes (which can be found from equation (\ref{schroedingerlike11234})) are in bijective correspondence to the zero modes of the operator $\mathcal{D}_{K}$. The same argument applies for the operator $\mathcal{D}_{K}^{\dag}$, with the difference that in this case the quasinormal modes are the complex conjugates of the previous case.

\noindent Making use of the operators $\mathcal{D}_{K}$ and $\mathcal{D}_{K}^{\dag}$ we can form an $N=2$ SUSY QM algebra. The supercharges of this algebra, $\mathcal{Q}_{K}$ and $\mathcal{Q}_{K}^{\dag}$ are defined in terms of  $\mathcal{D}_{K}$ and $\mathcal{D}_{K}^{\dag}$,
\begin{equation}\label{wit2}
\mathcal{Q}_{K}=\bigg{(}\begin{array}{ccc}
  0 & \mathcal{D}_{K} \\
  0 & 0  \\
\end{array}\bigg{)}, {\,}{\,}{\,}{\,}{\,}\mathcal{Q}^{\dag}_{K}=\bigg{(}\begin{array}{ccc}
  0 & 0 \\
  \mathcal{D}_{K}^{\dag} & 0  \\
\end{array}\bigg{)}.
\end{equation}
Additionally, the quantum Hamiltonian can be cast in following diagonal form,
\begin{equation}\label{wit4354}
\mathcal{H}_{K}=\bigg{(}\begin{array}{ccc}
  \mathcal{D}_{K}\mathcal{D}_{K}^{\dag} & 0 \\
  0 & \mathcal{D}_{K}^{\dag}\mathcal{D}_{K}  \\
\end{array}\bigg{)}
.\end{equation}
These operators, corresponding to the the radial part of
fermionic black hole system, are elements of an unbroken $N=2$ SUSY QM algebra, as we now demonstrate. The operators (\ref{wit2}) and (\ref{wit4354}), satisfy the $d=1$ SUSY algebra:
\begin{equation}\label{relationsforsusy}
\{\mathcal{Q}_{K},\mathcal{Q}^{\dag}_{K}\}=\mathcal{H}_{K}{\,}{\,},\mathcal{Q}_{K}^2=0,{\,}{\,}{\mathcal{Q}_{K}^{\dag}}^2=0
.\end{equation}
The Hilbert space of the supersymmetric quantum mechanical system, which we denote $\mathcal{H}$, is a $Z_2$ graded vector space, with the grading provided by the operator $\mathcal{W}$, the so-called Witten parity. The latter is an involution operator that commutes with the total Hamiltonian,
\begin{equation}\label{s45}
[\mathcal{W},\mathcal{H}_{K}]=0,
\end{equation}
and also, anti-commutes with the supercharges,
\begin{equation}\label{s5}
\{\mathcal{W},\mathcal{Q}_{K}\}=\{\mathcal{W},\mathcal{Q}_{K}^{\dag}\}=0.
\end{equation}
In addition, the operator $\mathcal{W}$ being a projection operator, satisfies,
\begin{equation}\label{s6}
\mathcal{W}^{2}=1.
\end{equation}
The Witten parity $\mathcal{W}$, spans the total Hilbert space into equivalent $Z_2$ subspaces. Therefore, the total Hilbert space of the quantum system is written:
\begin{equation}\label{fgjhil}
\mathcal{H}=\mathcal{H}^+\oplus \mathcal{H}^-,
\end{equation}
with the vectors corresponding to the two subspaces $\mathcal{H}^{\pm}$, classified to even and odd parity states, according to their Witten parity:
\begin{equation}\label{shoes}
\mathcal{H}^{\pm}=\mathcal{P}^{\pm}\mathcal{H}=\{|\psi\rangle :
\mathcal{W}|\psi\rangle=\pm |\psi\rangle \}.
\end{equation}
In addition, the corresponding Hamiltonians of the $Z_2$ graded spaces are:
\begin{equation}\label{h1}
{\mathcal{H}}_{+}=\mathcal{D}_{K}{\,}\mathcal{D}_{K}^{\dag},{\,}{\,}{\,}{\,}{\,}{\,}{\,}{\mathcal{H}}_{-}=\mathcal{D}_{K}^{\dag}{\,}\mathcal{D}_{K}.
\end{equation}
In the present case, the operator $\mathcal{W}$, can be represented in the following form:
\begin{equation}\label{wittndrf}
\mathcal{W}=\bigg{(}\begin{array}{ccc}
  1 & 0 \\
  0 & -1  \\
\end{array}\bigg{)}.
\end{equation}
In equation (\ref{shoes}) the operator $\mathcal{P}$, is defined in such a way, so that the eigenstates of $\mathcal{P}^{\pm}$, which are, $|\psi^{\pm}\rangle$, satisfy:
\begin{equation}\label{fd1}
P^{\pm}|\psi^{\pm}\rangle =\pm |\psi^{\pm}\rangle.
\end{equation}
We call them positive and negative parity eigenstates. Using the representation (\ref{wittndrf}) for the Witten parity operator,
the parity eigenstates are represented by,
\begin{equation}\label{phi5}
|\psi^{+}\rangle =\left(%
\begin{array}{c}
  |\phi^{+}\rangle \\
  0 \\
\end{array}%
\right),{\,}{\,}{\,}
|\psi^{-}\rangle =\left(%
\begin{array}{c}
  0 \\
  |\phi^{-}\rangle \\
\end{array}%
\right),
\end{equation}
with $|\psi^{\pm}\rangle$ $\epsilon$ $\mathcal{H}^{\pm}$. Turning back to the fermionic system at hand, we write the fermionic states of the system (\ref{ait34e11}) and (\ref{aigfgfhte21}) in terms of the SUSY QM algebra, that is:
\begin{equation}\label{fdgdfgh}
|\phi^{-}\rangle=|\phi ^{-}_K \rangle =\left(%
\begin{array}{c}
  R'_{-\frac{1}{2}}(r) \\
  R_{+\frac{1}{2}}(r) \\
\end{array}%
\right),{\,}{\,}{\,}|\phi^{+}\rangle=|\phi^{+}_K\rangle= \left(%
\begin{array}{c}
   \Big{(}R'_{-\frac{1}{2}}(r)\Big{)}^* \\
 \Big{(}R_{+\frac{1}{2}}(r)\Big{)}^* \\
\end{array}%
\right).
\end{equation}
Hence, the corresponding even and odd parity SUSY quantum states $|\psi^{+}\rangle$ and $|\psi^{-}\rangle$, are written in terms of $|\phi ^{-}_K \rangle$ and $|\phi ^{+}_K \rangle$:
\begin{equation}\label{phi5}
|\psi^{+}\rangle =\left(%
\begin{array}{c}
  |\phi ^{+}_K \rangle \\
  0 \\
\end{array}%
\right),{\,}{\,}{\,}
|\psi^{-}\rangle =\left(%
\begin{array}{c}
  0 \\
  |\phi ^{-}_K \rangle \\
\end{array}%
\right).
\end{equation}
When Fredholm operators are used, supersymmetry is considered unbroken if the Witten index is a non-zero integer. In this paper we shall not make use of Fredholm operators. Therefore, we shall need a generalization of the Fredholm index (and of the corresponding Witten index).
The heat-kernel regularized index, both for the operator $A$, 
that is $\mathrm{ind}_tA$ and for the Witten index, $\Delta_t$,
is defined as follows \cite{thaller,susyqm}:
\begin{align}\label{heatkerw}
& \mathrm{ind}_tA=\mathrm{Tr}(-\mathcal{W}e^{-tA^{\dag}A})=\mathrm{tr}_{-}(-\mathcal{W}e^{-tA^{\dag}A})-\mathrm{tr}_{+}(-\mathcal{W}e^{-tAA^{\dag}}) \\ \notag 
& \Delta_t=\lim_{t\rightarrow \infty}\mathrm{ind}_tA.
\end{align}
In the above, $t>0$, and  additionally the $\mathrm{tr}_{\pm }$ stands for the trace in the subspaces $\mathcal{H}^{\pm}$. The heat-kernel regularized index is defined for operators that are trace class (in our case, the operator $\mathrm{tr}(-\mathcal{W}e^{-tA^{\dag}A})$ must be trace class), that is, they have a finite trace norm. This is independent of the orthonormal basis describing the Hilbert space. From the Banach space of all trace class operators, we shall be interested in the subspace spanned by the $A$ and $A^{\dag}$ and their product $A{\,}A^{\dag}$. We now turn our focus on the regularized Witten index corresponding to the case at hand.

\noindent The equations of the quasinormal modes, $\mathcal{D}_K| \phi^{-}\rangle =0$ and it's conjugate
$\mathcal{D}_{K}^{\dag}|\phi^{+}\rangle =0$ have complex conjugate solutions. Obviously we have a bijective correspondence between the quasinormal modes, given by equation $\mathcal{D}_{K}|\phi^{-}\rangle =0$, and their complex conjugate counterparts, given by $\mathcal{D}_{K}^{\dag}|\phi^{+}\rangle=0$. It is obvious that this bijective correspondence holds between the zero modes of the matrices $\mathcal{D}_K$ and $\mathcal{D}^{\dag}_K$. Therefore,
\begin{equation}\label{keeerrrr}
\mathrm{ker}\mathcal{D}_{K}=\mathrm{ker}\mathcal{D}_{K}^{\dag}\neq 0,
\end{equation}
which in turn implies
$\mathrm{ker}\mathcal{D}_{K}\mathcal{D}_{K}^{\dag}=\mathrm{ker}\mathcal{D}_{K}\mathcal{D}_{K}^{\dag}\neq 0$. As a consequence, the following relation holds for the operators
$e^{-t\mathcal{D}_{K}^{\dag}\mathcal{D}_{K}}$ and $e^{-t\mathcal{D}_{K}\mathcal{D}^{\dag}_{K}}$ 
\begin{equation}\label{qggeeee123}
\mathrm{tr}_{-}e^{-t\mathcal{D}_{K}^{\dag}\mathcal{D}_{K}}=\mathrm{tr}_{+}e^{-t\mathcal{D}_{K}\mathcal{D}_{K}^{\dag}}
.\end{equation}
Recall that $\mathrm{tr}_{\pm }$ stands for the trace in the subspaces $\mathcal{H}^{\pm}$. As a consequence of relation (\ref{qggeeee123}), the regularized index of the operator $\mathcal{D}_{K}$ is equal to zero. Consequently the regularized Witten index is also zero. Hence, since the relation (\ref{keeerrrr}) holds true, the fermionic system possesses an unbroken
$N=2$ SUSY QM algebra. 

\noindent Using the notation of relation (\ref{phi5}) the equation $\mathcal{D}_{K}|\phi^{-}\rangle=0$ has a direct representative equation for the supercharge, namely,
$Q|\psi^{-}_0\rangle=0$. This implies that the zero mode eigenstate $|\psi^{-}_0\rangle$
is a negative Witten parity eigenstate, and is equal to:
\begin{equation}\label{neg}
|\psi^{-}_0\rangle=\left(%
\begin{array}{c}
  0\\
  0\\
   R'_{-\frac{1}{2}}(r) \\
  R_{+\frac{1}{2}}(r) \\
\end{array}%
\right)
.\end{equation}
In the same vain, the positive Witten parity vacuum eigenstate is,
\begin{equation}\label{negkar}
|\psi^{+}_0\rangle =\left(
\begin{array}{c}
  \Big{(}R'_{-\frac{1}{2}}(r)\Big{)}^* \\
 \Big{(}R_{+\frac{1}{2}}(r)\Big{)}^* \\
   0 \\
   0 \\
\end{array}
\right)
.\end{equation}
The angular case can be treated accordingly, following the same line of argument, as in the radial part of the Dirac equation. Adopting the notation of the previous paragraphs, the algebra can be built on the matrices $\mathcal{D}_{R_{\theta}}$ and $\mathcal{D}_{K_{\theta}}^{\dag}$. The matrix $\mathcal{D}_{K_{\theta}}$ is defined to be:
\begin{equation}\label{susyqmrnthetagkjhkk}
\mathcal{D}_{K_{\theta}}=\left(%
\begin{array}{cc}
 \mathcal{L}_{\frac{1}{2}} & -\lambda
 \\ \lambda &  \mathcal{L}_{\frac{1}{2}}^{\dag} \\
\end{array}%
\right)
,\end{equation}
acting on 
\begin{equation}\label{aituliue3}
|\phi ^{-}_{K_{\theta}} \rangle=\left(%
\begin{array}{c}
   S_{\frac{1}{2}}(\theta)\\
 S_{-\frac{1}{2}}(\theta) \\
\end{array}%
\right).
\end{equation}
which is a direct consequence of equation (\ref{finalan111g}). It's conjugate equals to:
\begin{equation}\label{susydgfgmrntheta}
\mathcal{D}_{K_{\theta}}^{\dag}=\left(%
\begin{array}{cc}
 \mathcal{L}_{\frac{1}{2}}^* & \lambda
 \\ -\lambda & {\mathcal{L}_{\frac{1}{2}}^{\dag}}^*\\
\end{array}%
\right)
.\end{equation}
Recall that the operators $\mathcal{L}_{\frac{1}{2}}$ and $\mathcal{L}_{\frac{1}{2}}^{\dag}$ contain $\omega$, which is a complex number. We must note that the situation at hand is much more complicated in comparison to the angular case. It is obvious that, since quasinormal modes exist, the operator $\mathcal{D}_{K_{\theta}}$ certainly has a set of discrete zero modes, that belong to a countable set of complex numbers. However, we cannot argue that the same holds for the operator $\mathcal{D}_{K_{\theta}}^{\dag}$. In the radial case, we could solve the zero mode problem of the two corresponding operators simultaneously, since the zero modes of the operators were complex conjugate (a proof for existence of solutions in a much more general setup see \cite{english}). But in the angular case, we cannot use the same argument. Therefore, we can argue that in general, the number of zero modes of the two matrices are not equal. Hence, we could naively argue that supersymmetry is unbroken in this case, but for different reasons in comparison to the radial case. This naive argument is based on the fact that the operators have not the same number of zero modes, and therefore, supersymmetry is unbroken. This however would be true only in the case the operators were Fredholm, which are not (since $\mathrm{dim{\,}kerD_{k_{\theta}}}\rightarrow \infty$). Moreover, we cannot be sure whether the operator $\mathcal{D}_{K_{\theta}}^{\dag}$ is trace-class.  Therefore, we conclude that only the radial part of a Dirac fermionic system in the Kerr black hole background can be associated to a $N=2$ SUSY QM algebra. 

\noindent Before closing this section, we must note that up to date, the important theoretical issue that addresses the nature of neutrino, that is whether it is Dirac or Majorana, has not be answered successfully yet. Hence, any information on the effect of massless (if the neutrino can be considered massless) fermions in nature is invaluable. We studied a massless Dirac fermion in the most realistic curved four dimensional gravitational background and found an underlying supersymmetry. This supersymmetry is unbroken, a fact that is guaranteed by the existence of quasinormal modes of the fermion in the same background. It would certainly be interesting to study if supersymmetric structures exist when massive Dirac fermions and also when Majorana fermions are considered. Of course this would require a complete study of the quasinormal modes of massive Dirac fermions and of Majorana fermions in Kerr backgrounds.

\section{Fermion Quasinormal modes for Kerr-Newman, Reissner-N\"{o}rdstrom and Schwarzschild Black Holes}

\subsection{The Kerr-Newman Black Hole}

In this section we further explore whether supersymmetric structures underlie any other fermion systems in curved gravitational backgrounds. We shall study first the Kerr-Newman black hole, which is the only asymptotically flat solution of the Einstein equations, with electrifying vacuum. We consider a massive fermion of Dirac type in such a black hole background. Following the line of research of the previous section and adopting the notation of reference \cite{quasidirac6}, the fermionic equations of motion can be recast as,
\begin{align}\label{finalangKN}
&\mathcal{L}_{\frac{1}{2}}S_{\frac{1}{2}}=-(\lambda-am_F\cos\theta )
S_{-\frac{1}{2}}\\ \notag &
\mathcal{L}_{\frac{1}{2}}^{\dag}S_{-\frac{1}{2}}=(\lambda+am_F\cos\theta )
S_{\frac{1}{2}},
\end{align}
in reference to the angular part. The radial part can be written as:
\begin{align}\label{finalradKN}
&\sqrt{\Delta}\mathcal{D}_{0}R_{-\frac{1}{2}}=(\lambda+im_Fr) R'_{+\frac{1}{2}}\\
\notag &
\sqrt{\Delta}\mathcal{D}_{0}^{\dag}R'_{+\frac{1}{2}}=(\lambda-im_Fr)
\sqrt{2}R_{-\frac{1}{2}},
\end{align}
with $R'_{\frac{1}{2}}=\sqrt{\Delta}R_{+\frac{1}{2}}$ and $\Delta=r^2+a^2-2Mr+Q^2$. 
The operators that appear in relations (\ref{finalangKN}) and (\ref{finalradKN}) are equal to:
\begin{align}\label{basoper2KN}
&\mathcal{D}_n=\partial_r+i\frac{K}{\Delta}+\frac{n}{\Delta}\frac{\mathrm{d}\Delta}{\mathrm{d}r},
{\,}{\,}{\,}\mathcal{D}_n^{\dag}=\partial_r-i\frac{K}{\Delta}+\frac{n}{\Delta}\frac{\mathrm{d}\Delta}{\mathrm{d}r}\\
\notag & \mathcal{L}_n=\partial_{\theta}-a\omega\sin\theta+\frac{m}{\sin
\theta}+n\cot\theta,
{\,}{\,}{\,}\mathcal{L}_n^{\dag}=\partial_{\theta}+a\omega\sin\theta-\frac{m}{\sin
\theta}+n\cot\theta,
\end{align}
with $K=(r^2+a^2)\omega+am$. Note that ``$\lambda$'' is the same as in the Kerr case, since the angular equation can be reduced to Eq. (\ref{angularpareal}), corresponding to the Kerr case \cite{quasidirac6}. In the following paragraphs, we shall study both the massless and massive fermion case. As we shall see, the mass can introduce some complications to our initial arguments, but the final result is the same as in the massless case.

\subsubsection{Massless Fermion case}

Consider the radial part of a massless Dirac fermion first. As in the previous section, from the equations of motion (\ref{finalradKN}), we can construct the matrix $\mathcal{D}_R$, on which the supersymmetric quantum algebra can be built on. This matrix is defined to be:
\begin{equation}\label{susyqmrnm}
\mathcal{D}_R=\left(%
\begin{array}{cc}
 \sqrt{\Delta}D_0 & \lambda
 \\ \lambda & \sqrt{\Delta} D_0^{\dag}\\
\end{array}%
\right)
,\end{equation}
acting on the vector:
\begin{equation}\label{aite1}
\left(%
\begin{array}{c}
  R_{-\frac{1}{2}}(r) \\
  R'_{+\frac{1}{2}}(r) \\
\end{array}%
\right).
\end{equation}
Using the explicit form of the operators defined in equation (\ref{basoper2KN}), the operator $\mathcal{D}_R$ equals to:
\begin{equation}\label{susyqmrnm}
\mathcal{D}_R=\left(%
\begin{array}{cc}
 \sqrt{\Delta}\Big{(}\partial_r+i\frac{K}{\Delta}\Big{)} & \lambda
 \\ \lambda & \sqrt{\Delta} \Big{(}\partial_r-i\frac{K}{\Delta}\Big{)}\\
\end{array}%
\right)
.\end{equation}
We can easily obtain it's adjoint, $\mathcal{D}_R^{\dag}$, which is equal to:
\begin{equation}\label{susyqmrnmdag}
\mathcal{D}_R^{\dag}=\left(%
\begin{array}{cc}
 \sqrt{\Delta}\Big{(}\partial_r-i\frac{K^*}{\Delta}\Big{)} & \lambda
 \\ \lambda & \sqrt{\Delta} \Big{(}\partial_r+i\frac{K^*}{\Delta}\Big{)}\\
\end{array}%
\right)
,\end{equation}
acting on,
\begin{equation}\label{aite2}
\left(%
\begin{array}{c}
   \Big{(}R'_{+\frac{1}{2}}(r)\Big{)}^* \\
 \Big{(}R_{-\frac{1}{2}}(r)\Big{)}^* \\
\end{array}%
\right)
.\end{equation}
The number of the zero modes of the operator $\mathcal{D}_{R}^{\dag}$ are bijectively related to the number of the zero modes of $\mathcal{D}_{R}$. This is because, the set of the zero modes of $\mathcal{D}_{R}$ are in one-to-one correspondence to the quasinormal modes, corresponding to the equation (\ref{finalradKN}). In the same vain, the zero modes of $\mathcal{D}_{R}^{\dag}$ are in one-to-one correspondence to the quasinormal modes, corresponding to complex conjugate of the equation (\ref{finalradKN}). It is necessary to note that, in order to obtain consistent solutions for the complex conjugate of equation  (\ref{finalradKN}), the wave functions must obey the complex conjugate boundary conditions of the wave functions that correspond to equation (\ref{finalradKN}). The existence of a solution for this case is obvious, but can be further justified by a theorem on second order differential equations \cite{english}. We are not interested in the specific form of the solutions, but only in the fact that the zero modes of the operators  $\mathcal{D}_{R}^{\dag}$ and $\mathcal{D}_{R}$ have a bijective correspondence. The situation at hand is very similar to the massless Kerr fermion case of the previous section. Having found a correspondence between the zero modes, it is easy to verify that supersymmetry is unbroken, with the heat-kernel regularized Witten index being equal to zero. Let us see this in detail. The SUSY QM algebra can be built on the supercharges,
\begin{equation}\label{wit2kn}
\mathcal{Q}_R=\bigg{(}\begin{array}{ccc}
  0 & \mathcal{D}_{R} \\
  0 & 0  \\
\end{array}\bigg{)}, {\,}{\,}{\,}{\,}{\,}\mathcal{Q}^{\dag}_R=\bigg{(}\begin{array}{ccc}
  0 & 0 \\
  \mathcal{D}_{R}^{\dag} & 0  \\
\end{array}\bigg{)}
,\end{equation}
and also the corresponding Hamiltonian,
\begin{equation}\label{wit4}
\mathcal{H}_R=\bigg{(}\begin{array}{ccc}
  \mathcal{D}_{R}\mathcal{D}_{R}^{\dag} & 0 \\
  0 & \mathcal{D}_{R}^{\dag}\mathcal{D}_{R}  \\
\end{array}\bigg{)}
.\end{equation}
These satisfy $\{\mathcal{Q}_R,{\mathcal{Q}_R}^{\dag}\}=\mathcal{H}_R$,
$\mathcal{Q}^2_R=0$, ${\mathcal{Q}^{\dag}_R}^2=0$ and $[\mathcal{W},\mathcal{H}_R]=0$.
Hence, an $N=2$ SUSY QM algebra underlies the radial part of
fermionic Kerr-Newman black hole system. In addition, as a result of the bijective correspondence of the two matrices zero modes, we have,
\begin{equation}\label{keeerrrrnewm}
\mathrm{ker}\mathcal{D}_R=\mathrm{ker}\mathcal{D}^{\dag}_R\neq 0
,\end{equation}
which in turn implies,
$\mathrm{ker}\mathcal{D}_R\mathcal{D}^{\dag}_R=\mathrm{ker}\mathcal{D}_R\mathcal{D}^{\dag}_R \neq 0$. As a consequence, the following relation holds for the operators
$e^{-t\mathcal{D}^{\dag}_R\mathcal{D}_R}$ and $e^{-t\mathcal{D}_R\mathcal{D}^{\dag}_R}$ 
\begin{equation}\label{qggeeee1}
\mathrm{tr}_{-}e^{-t\mathcal{D}^{\dag}_R\mathcal{D}_R}=\mathrm{tr}_{+}e^{-t\mathcal{D}_R\mathcal{D}^{\dag}_R}
.\end{equation}
The above result holds only in the case the operator $\mathrm{tr}(-\mathcal{W}e^{\mathcal{D}_R\mathcal{D}^{\dag}_R})$ is trace class, which is true, since the operators $\mathcal{D}_R$, $\mathcal{D}^{\dag}_R$ and consequently the operators $\mathcal{D}_R\mathcal{D}^{\dag}_R$, $\mathcal{D}^{\dag}_R\mathcal{D}_R$ are trace class \cite{thaller}. Hence, the regularized Witten index is zero, and this fact, in conjunction with equation (\ref{keeerrrrnewm}), implies that the radial part of the fermionic system has unbroken supersymmetry. We omit the study of the angular part of the fermionic system since, as in the Kerr case, supersymmetry is broken, for the same reasons.

\subsubsection{A brief Discussion}

We mentioned in the above that in order relation (\ref{qggeeee1}) holds true, the operator $\mathrm{tr}(-\mathcal{W}e^{\mathcal{D}_R\mathcal{D}^{\dag}_R})$ must be trace class. We argued that this is true since the operators $\mathcal{D}_R{\,}\mathcal{D}_R^{\dag}$ and $\mathcal{D}_R^{\dag}{\,}\mathcal{D}_R$ are trace class. However, this is not prerequisite for our case, and the only requirement for $\mathrm{tr}(-\mathcal{W}e^{\mathcal{D}_R\mathcal{D}^{\dag}_R})$ to be trace class is that $\mathcal{D}_R{\,}\mathcal{D}_R^{\dag}-\mathcal{D}_R^{\dag}{\,}\mathcal{D}_R$ is trace class. Indeed, by virtue of a theorem (see for example \cite{thaller}, page 161, comments before Theorem 5.20 and also Theorem 5.22), if $T$ and $S$ are two self-adjoint operators, and $T-S$ is trace class, then $f(T)-f(S)$ is also trace class. The function $f$ is a map $f: \mathcal{R}\rightarrow \mathcal{R}$, satisfying:
\begin{equation}\label{ezeltag}
\Big{|} \int_{\infty}^{\infty}\hat{f}(p)(1+| p | ) \mathrm{d}p \Big{|} < \infty
,\end{equation}
with $\hat{f}$ the Fourier transform of $f$. It is clear that in our case, the function $f$ is the trace of the exponential of the operators $\mathcal{D}_R{\,}\mathcal{D}_R^{\dag}$ and $\mathcal{D}_R^{\dag}{\,}\mathcal{D}_R$.  Hence, the theorem applies perfectly in the present situation.

\subsubsection{Massive Fermion case}

We extend the study in the case the Dirac fermion has mass $m_F$. From the equations of motion (\ref{finalradKN}) we can construct the matrix,
\begin{equation}\label{susyqmrnmassive}
\mathcal{D}_{MR}=\left(%
\begin{array}{cc}
 \sqrt{\Delta}\Big{(}\partial_r+i\frac{K}{\Delta}\Big{)}& \lambda+im_Fr
 \\ \lambda-im_Fr &  \sqrt{\Delta}\Big{(}\partial_r-i\frac{K}{\Delta}\Big{)}\\
\end{array}%
\right)
,\end{equation}
which acts on the vector: 
\begin{equation}\label{massaite2}
\left(%
\begin{array}{c}
   R'_{+\frac{1}{2}}(r) \\
 R_{-\frac{1}{2}}(r) \\
\end{array}%
\right).
\end{equation}
The adjoint of the matrix $\mathcal{D}_{MR}$ is equal to:
\begin{equation}\label{susyqmrnmassive}
\mathcal{D}_{MR}^{\dag}=\left(%
\begin{array}{cc}
\sqrt{\Delta}\Big{(}\partial_r-i\frac{K^*}{\Delta}\Big{)}& \lambda+im_Fr
 \\ \lambda-im_Fr &  \sqrt{\Delta}\Big{(}\partial_r+i\frac{K^*}{\Delta}\Big{)}\\
\end{array}%
\right)
.\end{equation}
Unlike the massless case, it is not obvious if the zero modes of $\mathcal{D}_{MR}^{\dag}$ exist at all. One must solve the equation $\mathcal{D}_{MR}^{\dag}\psi =0$, subject to the complex conjugate boundary conditions corresponding to equation (\ref{finalradKN}). The zero modes of the matrix $\mathcal{D}_{MR}$ exist, and the countable set that these belong to is in one-to-one correspondence to the quasinormal modes for a massive fermion in the Kerr-Newman background. The latter were studied in detail in \cite{quasidirac6}. We are thus confronted with a non-trivial problem. In any case, there are two alternative situations that can occur, in reference to the zero modes of $\mathcal{D}_{MR}^{\dag}$. Either $\mathrm{dim{\,}ker}\mathcal{D}_{MR}^{\dag}=0$ or $\mathrm{dim{\,} ker}\mathcal{D}_{MR}^{\dag}\neq 0$, and we cannot be sure which case is true unless we solve explicitly the equation $\mathcal{D}_{MR}^{\dag}\psi =0$. Nonetheless, we can answer whether supersymmetry is broken or not if we make use of a theorem. First note that, we can write $\mathcal{D}_{MR}=\mathcal{D}_{R}+C$, with $C$ the odd symmetric matrix \cite{thaller},
\begin{equation}\label{susyqmrnmassive}
C=\left(%
\begin{array}{cc}
 0 & im_Fr
 \\ -im_Fr & 0\\
\end{array}%
\right)
,\end{equation}
and also, the $\mathcal{D}_R$ is the massless Kerr-Newman case matrix defined in equation (\ref{susyqmrnm}).
Now, if the operator $\mathrm{tr}\mathcal{W}e^{-t(\mathcal{D}+C)^2}$ is trace class, the following theorem holds (see for example \cite{thaller} page 168, Theorem 5.28),
\begin{equation}\label{indperturbatrn1}
\mathrm{ind}_{t}(\mathcal{D}+C)=\mathrm{ind}_{t}\mathcal{D}
,\end{equation}
with $C$ a symmetric odd operator and $\mathcal{D}$ any trace class operator. That is, the regularized index of the operator $\mathcal{D}+C$ is equal to the index of the operator $\mathcal{D}$. In our case, this reads, $\mathrm{ind}_{t}(\mathcal{D}_R+C)=\mathrm{ind}_{t}\mathcal{D}_R$. Recall that the Witten index and the heat kernel regularized index of $\mathcal{D}_R$ in the Kerr-Newman background are both zero, hence $\mathrm{ind}_{t}\mathcal{D}_R=0$. By virtue of theorem (\ref{indperturbatrn1}), $\mathrm{ind}_{t}(\mathcal{D}_R+C)=0$. Hence, in conjunction with the fact that $\mathrm{ker}\mathcal{D}_R\neq 0$, we may argue that supersymmetry is unbroken even for the massive fermion case.

\subsection{The Reissner-N\"{o}rdstrom and Schwarzschild Black Hole }

Following the lines of argument of the previous sections, it can be easily proven that an unbroken $N=2$ SUSY QM algebra underlies the radial part of massless and massive Dirac fermionic systems in Reissner-N\"{o}rdstrom and Schwarzschild black hole backgrounds. For a detailed study of the corresponding fermionic quasinormal modes can be found in references \cite{quasidirac5} and \cite{quasidirac2}, respectively. We shall not get into details, since the results are identical to those of the Kerr-Newman case. Indeed, the radial part of the Dirac equation in the Reissner-N\"{o}rdstrom background can be reduced to the same form of equations, like Equation (\ref{finalradKN}). In the Reissner-N\"{o}rdstrom black hole, $\Delta=r^2-2Mr+Q^2$ and $\lambda^2=(j+\frac{1}{2})^2$, with $j$ positive. The charges of the supersymmetric quantum mechanical algebra are identical to those of the Kerr-Newmann case, with the appropriate replacement of $\Delta$ and $\lambda$. In the case of the Schwarzschild black hole the same arguments hold, but we should replace everywhere $\Delta=r^2-2Mr$ and $\lambda^2=(j+\frac{1}{2})^2$.

\section{The D-dimensional de Sitter Spacetime, Kerr-Newman-de Sitter Black Hole and Reissner-Nordstr\"{o}m-anti-de Sitter Spacetime}

In this section we shall  study whether the supersymmetric structures we found in the previous sections, also underlie de Sitter  and anti-de Sitter related spacetimes. We start first with a D-dimensional de Sitter spacetime, in which we consider a massive Dirac fermion. We adopt the notation of reference \cite{quasidirac1}. See also references \cite{quasidirac7,quasidirac3,quasidirac4}. In general, the metric of an $D$-dimensional spherically symmetric spacetime, is of the form,
\begin{equation}\label{ddesitter}
\mathrm{d}s^2=W(y)^2\mathrm{d}t^2-\frac{W(y)^2}{U(y)^2}\mathrm{d}y^2-\frac{W(y)^2}{V(y)^2}y^2\mathrm{d}\Omega^2_{D-2}
,\end{equation}
with $\mathrm{d}\Omega^2_{D-2}$, the metric of the $D-2$ dimensional unit sphere. In this spacetime, the Dirac equation of a fermion with mass $m_f$ can be cast as:
\begin{equation}\label{diracdessitteeer}
\left(%
\begin{array}{cc}
 m_f{\,}W& U{\,}\frac{\mathrm{d}}{\mathrm{d}y}+k\frac{V}{y}
 \\-U{\,}\frac{\mathrm{d}}{\mathrm{d}y}+k\frac{V}{y} & -m_f{\,}W\\
\end{array}%
\right)  \left(
\begin{array}{c}
   f^{(1)}_{\omega,k}(y)\\
  \\f^{(2)}_{\omega,k}(y)
\end{array}%
\right)=\omega \left(%
\begin{array}{c}
   f^{(1)}_{\omega,k}(y)\\
  \\f^{(2)}_{\omega,k}(y)
\end{array}%
\right)
,\end{equation}
with $k=\pm(\frac{D-2}{2}+n)$. In the D-dimensional de Sitter spacetime background, the quantities $U$, $V$ and $W$ are equal to:
\begin{equation}\label{akdfkffd}
U=1{\,}{\,}{\,}{\,}{\,}\frac{V}{y}=\frac{1}{L\sinh (\frac{y}{L})}{\,}{\,}{\,}{\,}{\,}W=\frac{1}{L\cosh (\frac{y}{L})}
.\end{equation}
In the above equation, $L$ is related to the cosmological constant. Using the coordinate $z=\tanh (\frac{y}{L})$, with $r=zL$, the Dirac equation (\ref{diracdessitteeer}) can be written:
\begin{align}\label{desitterdiracequation}
& \Big{[}\sqrt{1-z^2}\frac{\mathrm{d}}{\mathrm{d}{\,}z}-\frac{i\tilde{\omega }}{\sqrt{1-z^2}}\Big{]}R_1(y)-\Big{[}\frac{k}{z}-i\tilde{M}\Big{]}R_2(y)=0 \\ \notag & \Big{[}\sqrt{1-z^2}\frac{\mathrm{d}}{\mathrm{d}{\,}z}+\frac{i\tilde{\omega }}{\sqrt{1-z^2}}\Big{]}R_2(y)-\Big{[}\frac{k}{z}+i\tilde{M}\Big{]}R_1(y)=0,
\end{align}
with $\tilde{\omega }=\omega {\,}L$ and $\tilde{M}=m_f{\,}L$ and additionally,
\begin{equation}\label{functions}
R_1(y)=-i{\,}f^{(1)}_{\omega,k}(y)+f^{(2)}_{\omega,k}(y),{\,}{\,}{\,}{\,}{\,}{\,}{\,}{\,}{\,}{\,}R_2(y)=-i{\,}f^{(1)}_{\omega,k}(y)-f^{(2)}_{\omega,k}(y)
.\end{equation}
In the case $\tilde{M}=0$, equation (\ref{desitterdiracequation}) becomes:
\begin{align}\label{desitterdiracequatiofghjhn}
& \Big{[}\sqrt{1-z^2}\frac{\mathrm{d}}{\mathrm{d}{\,}z}-\frac{i\tilde{\omega }}{\sqrt{1-z^2}}\Big{]}R_1(y)-\frac{k}{z}R_2(y)=0 \\ \notag & \Big{[}\sqrt{1-z^2}\frac{\mathrm{d}}{\mathrm{d}{\,}z}+\frac{i\tilde{\omega }}{\sqrt{1-z^2}}\Big{]}R_2(y)-\frac{k}{z}R_1(y)=0.
\end{align}
The above equation gives the quasinormal mode spectrum of the fermionic system in this de-Sitter background. It is obvious that, also in this case, the quasinormal spectrum is in bijective correspondence to the zero modes of the operator:
\begin{equation}\label{susyqmrtyitin567m}
\mathcal{D}_{dS}=\left(%
\begin{array}{cc}
 \sqrt{1-z^2}\frac{\mathrm{d}}{\mathrm{d}{\,}z}-\frac{i\tilde{\omega }}{\sqrt{1-z^2}} & \frac{k}{z}
 \\ \frac{k}{z} & \sqrt{1-z^2}\frac{\mathrm{d}}{\mathrm{d}{\,}z}+\frac{i\tilde{\omega }}{\sqrt{1-z^2}}\\
\end{array}%
\right)
,\end{equation}
acting on the vector:
\begin{equation}\label{ait34ityitye1}
|\phi ^{-}_{dS} \rangle =\left(%
\begin{array}{c}
  R_1(y) \\
  R_2(y) \\
\end{array}%
\right)
,\end{equation}
and also  $\mathcal{D}_{dS}^{\dag}$:
\begin{equation}\label{susyqmtityirnmd2ag1}
\mathcal{D}_{dS}^{\dag}=\left(%
\begin{array}{cc}
 \sqrt{1-z^2}\frac{\mathrm{d}}{\mathrm{d}{\,}z}+\frac{i\tilde{\omega }}{\sqrt{1-z^2}} & \frac{k}{z}
 \\ \frac{k}{z} & \sqrt{1-z^2}\frac{\mathrm{d}}{\mathrm{d}{\,}z}-\frac{i\tilde{\omega }}{\sqrt{1-z^2}}\\
\end{array}%
\right)
,\end{equation}
acting on 
\begin{equation}\label{aigfgftyityihte2}
|\phi^{+}_{dS}\rangle =\left(%
\begin{array}{c}
   \Big{(} R_2(y)\Big{)}^* \\
 \Big{(}R_1(y)\Big{)}^* \\
\end{array}%
\right).
\end{equation}
As in the Kerr case, the operators can be used to form an unbroken $N=2$ SUSY QM algebra.
Without getting into details, the supercharges of this algebra, $\mathcal{Q}_{dS}$ and $\mathcal{Q}^{\dag}_{dS}$ are,
\begin{equation}\label{wittyutyi2}
\mathcal{Q}_{dS}=\bigg{(}\begin{array}{ccc}
  0 & \mathcal{D}_{dS} \\
  0 & 0  \\
\end{array}\bigg{)}, {\,}{\,}{\,}{\,}{\,}\mathcal{Q}_{dS}^{\dag}=\bigg{(}\begin{array}{ccc}
  0 & 0 \\
  \mathcal{D}_{dS}^{\dag} & 0  \\
\end{array}\bigg{)}
.\end{equation}
Moreover, the quantum Hamiltonian is,
\begin{equation}\label{witurutyu4}
\mathcal{H}_{dS}=\bigg{(}\begin{array}{ccc}
  \mathcal{D}_{dS}\mathcal{D}_{dS}^{\dag} & 0 \\
  0 & \mathcal{D}_{dS}^{\dag}\mathcal{D}_{dS}  \\
\end{array}\bigg{)}
.\end{equation}
Since, \begin{equation}\label{keeerrrrthrt}
\mathrm{ker}\mathcal{D}_{dS}=\mathrm{ker}\mathcal{D}_{dS}^{\dag}
,\end{equation}
which in turn implies,
$\mathrm{ker}\mathcal{D}_{dS}\mathcal{D}_{dS}^{\dag}=\mathrm{ker}\mathcal{D}_{dS}\mathcal{D}_{dS}^{\dag}$. As a consequence, the following relation holds for the operators,
$e^{-tD^{\dag}_{dS}\mathcal{D}_{dS}}$ and $e^{-t\mathcal{D}_{dS}\mathcal{D}_{dS}^{\dag}}$, 
\begin{equation}\label{qggeeee}
\mathrm{tr}_{-}e^{-t\mathcal{D}_{dS}^{\dag}\mathcal{D}_{dS}}=\mathrm{tr}_{+}e^{-t\mathcal{D}_{dS}\mathcal{D}_{dS}^{\dag}}.
\end{equation}  
Hence, supersymmetry is unbroken, just in the Kerr massless fermion case. The difference between the two problems however is that, in the de Sitter case, the whole system possesses this supersymmetry and in the Kerr case, only the radial part has this symmetry. The massive case can be also treated identically to the massive case of the Kerr-Newman black hole, yielding the result that the system possesses an unbroken $N=2$ SUSY QM. Before closing this section, we mention that the system of a Dirac fermionic field in the Kerr-Newman-de Sitter black hole background has also an $N=2$ SUSY QM, but we omit such a study since it is identical to the other cases we studied, with different supercharges and Hamiltonian of course. For a study of the fermionic quasinormal modes in a Kerr-Newman-de Sitter background, see reference \cite{quasidirac7}.

\subsection*{A Brief Presentation of the Reissner-Nordstr\"{o}m-anti-de Sitter Spacetime}

An interesting situation, different from the ones we met in the previous sections, occurs in the case a Dirac fermion is considered in a Reissner-Nordstr\"{o}m-anti-de Sitter spacetime. Particularly, as was established in reference \cite{oikonomoucolor}, this fermionic gravitational system has two unbroken $N=2$ SUSY QM algebras. We shall briefly present it since we are going to use it in the following (for a detailed analysis see \cite{oikonomoucolor}). The metric in a $D$-dimensional Reissner-Nordstr\"{o}m-anti-de Sitter spacetime is
given by:
\begin{equation}\label{RNmetric}
\mathrm{d}s^2=-f(r)\mathrm{d}t^2+\frac{1}{f(r)}\mathrm{d}r^2+r^2\mathrm{d}\Omega^2_{D-2,k}
,\end{equation}
where $f(r)$ is equal to:
\begin{equation}\label{fr}
f(r)=k+\frac{r^2}{L^2}+\frac{Q^2}{4r^{2D-6}}-\Big{(}\frac{r_0}{r}\Big{)}^{d-3}
.\end{equation}
In the above equation, $L$ is the AdS radius, Q is the black hole
charge, and $r_0$ is related to the black hole mass $M$. The
$\mathrm{d}\Omega^2_{D-2,k}$ is the metric of constant curvature,
with $k$ characterizing the curvature. The value $k>0$
characterizes the metric of an $D-2$ dimensional sphere, while the
$k=0$ describes $R^{D-2}$. Finally, when $k<0$, it describes
$H^{D-2}$. In the 4-dimensional case, the $k=0$
Reissner-Nordstr\"{o}m-anti-de Sitter metric is,
\begin{equation}\label{metric4}
\mathrm{d}s^2=-f(r)\mathrm{t}^2+\frac{1}{f(r)}\mathrm{d}r^2+r^2(\mathrm{d}x^2+\mathrm{d}y^2)
.\end{equation}
As is demonstrated in \cite{oikonomoucolor}, the first unbroken $N=2$ SUSY QM algebra, which we denote as $\mathcal{N}_1$, can be constructed by using the following operator:
\begin{equation}\label{susyqmrn567m12}
{{{{{\mathcal{D}}_{RN}}}}}=\left(%
\begin{array}{cc}
 \partial_r-\frac{i}{f}(\omega+qA_t) & \frac{i}{r\sqrt{f}}k_x
 \\  -\frac{i}{r\sqrt{f}}k_x & \partial_r+\frac{i}{f}(\omega+qA_t)\\
\end{array}%
\right)
.\end{equation}
The supercharges of the $\mathcal{N}_1$ SUSY algebra, which we denote ${\mathcal{Q}}_{RN}$
and ${\mathcal{Q}}^{\dag}_{RN}$,  are equal to:
\begin{equation}\label{wit2dr}
{\mathcal{Q}}_{RN}=\bigg{(}\begin{array}{ccc}
  0 & {{{{{\mathcal{D}}_{RN}}}}} \\
  0 & 0  \\
\end{array}\bigg{)}, {\,}{\,}{\,}{\,}{\,}{\mathcal{Q}}^{\dag}_{RN}=\bigg{(}\begin{array}{ccc}
  0 & 0 \\
  {{{{{\mathcal{D}}_{RN}}}}}^{\dag} & 0  \\
\end{array}\bigg{)}
.\end{equation}
Moreover, the quantum Hamiltonian is equal to,
\begin{equation}\label{wit4354dr}
H_{RN}=\bigg{(}\begin{array}{ccc}
  {{{{\mathcal{D}}_{RN}}}}{{{{\mathcal{D}}_{RN}}}}^{\dag} & 0 \\
  0 & {{{{\mathcal{D}}_{RN}}}}^{\dag}{{{{\mathcal{D}}_{RN}}}}  \\
\end{array}\bigg{)}
.\end{equation}
The second unbroken $N=2$ SUSY QM algebra, denoted as $\mathcal{N}_2$, can be built on the matrix:
\begin{equation}\label{susyqmrn567m}
{\mathcal{D}}_{RN'}=\left(%
\begin{array}{cc}
 \partial_r-\frac{i}{f}(-\omega-qA_t) & -\frac{i}{r\sqrt{f}}k_x
 \\  \frac{i}{r\sqrt{f}}k_x & \partial_r+\frac{i}{f}(-\omega-qA_t)\\
\end{array}%
\right)
,\end{equation}
with the corresponding supercharges,
\begin{equation}\label{wit2dr1}
{\mathcal{Q}}_{RN'}=\bigg{(}\begin{array}{ccc}
  0 & {\mathcal{D}}_{RN'} \\
  0 & 0  \\
\end{array}\bigg{)}, {\,}{\,}{\,}{\,}{\,}{\mathcal{Q}}^{\dag}_{RN'}=\bigg{(}\begin{array}{ccc}
  0 & 0 \\
  {\mathcal{D}}_{RN'}^{\dag} & 0  \\
\end{array}\bigg{)}
.\end{equation}
The Hamiltonian of $\mathcal{N}_2$ is
\begin{equation}\label{wit4354dr1231}
H_{RN'}=\bigg{(}\begin{array}{ccc}
  {\mathcal{D}}_{RN'}{\mathcal{D}}_{RN'}^{\dag} & 0 \\
  0 & {\mathcal{D}}_{RN'}^{\dag}{\mathcal{D}}_{RN'}  \\
\end{array}\bigg{)}
.\end{equation}
Hence the Dirac
fermionic system in an Reissner-Nordstr\"{o}m-anti-de Sitter
background, possesses an supersymmetric structure that is the direct sum
of two $N=2$ supersymmetries, namely:
\begin{equation}\label{directsum}
N_{total}=\mathcal{N}_1\oplus \mathcal{N}_2
.\end{equation}

\section{Physical and Geometric Implications of the SUSY QM Algebra}

\subsection{Extended Supersymmetric-Higher Representation Algebras}

As we have demonstrated, in the case of the Reissner-Nordstr\"{o}m-anti-de Sitter gravitational fermionic system, there are two $N=2$ SUSY QM algebras. In view of this fact, a natural question that springs to mind, is whether the SUSY QM algebras that underlie the gravitational systems have an extended (with $N=4,6...$) one dimensional supersymmetry origin. This argument is further supported from the fact that \cite{gibbons} non-Abelian uplifted magnetic dilatonic black holes are connected to a $N=4$ SUSY QM algebra. It is obvious that the theoretical frameworks are different in the two cases, but nevertheless the underlying SUSY QM structure motivates us to search for an extended supersymmetry structure.

\noindent Unfortunately, there is no extended SUSY QM algebra underlying the gravitational systems. In fact, the extended SUSY QM algebra is always connected to a global spacetime supersymmetry in four dimensions, and clearly there is no such structure in the gravitational systems we studied. The only case for which there exists a more rich SUSY QM structure than the $N=2$ SUSY QM, is in the case of the Reissner-Nordstr\"{o}m-anti-de Sitter gravitational fermionic system. In that case, the two $N=2$ supersymmetries can be combined to a higher representation of a single $N=2$, $d=1$ supersymmetry. Indeed, the supercharges of this representation, which we denote ${\mathcal{Q}}_{T}$ and  ${\mathcal{Q}}_{T}^{\dag}$ are equal to:
\begin{equation}\label{connectirtyrtons}
{\mathcal{Q}}_{T}= \left ( \begin{array}{cccc}
  0 & 0 & 0 & 0 \\
  {\mathcal{D}}_{RN} & 0 & 0 & 0 \\
0 & 0 & 0 & 0 \\
0 & 0 & {\mathcal{D}}_{RN'}^{\dag} & 0  \\
\end{array} \right),{\,}{\,}{\,}{\,}{\mathcal{Q}}_{T}^{\dag}= \left ( \begin{array}{cccc}
  0 &  {\mathcal{D}}_{RN}^{\dag} & 0 & 0 \\
  0 & 0 & 0 & 0 \\
0 & 0 & 0 & {\mathcal{D}}_{RN'} \\
0 & 0 & 0 & 0  \\
\end{array} \right)
.\end{equation}
Accordingly, the Hamiltonian of the combined quantum system, which we denote $H_T$, reads,
\begin{equation}\label{connections1dtr}
H_T= \left ( \begin{array}{cccc}
  {\mathcal{D}}_{RN}^{\dag}{\mathcal{D}}_{RN} & 0 & 0 & 0 \\
  0 & {\mathcal{D}}_{RN}{\mathcal{D}}_{RN}^{\dag} & 0 & 0 \\
0 & 0 & {\mathcal{D}}_{RN'}{\mathcal{D}}_{RN'}^{\dag} & 0 \\
0 & 0 & 0 & {\mathcal{D}}_{RN'}^{\dag}{\mathcal{D}}_{RN'}  \\
\end{array} \right)
.\end{equation}
The operators (\ref{connectirtyrtons}) and (\ref{connections1dtr}), satisfy the $N=2$, $d=1$ SUSY QM algebra, namely:
\begin{equation}\label{mousikisimagne}
\{ {\mathcal{Q}}_{T},{\mathcal{Q}}_{T}^{\dag}\}=H_{T},{\,}{\,}{\mathcal{Q}}_{T}^2=0,{\,}{\,}{{\mathcal{Q}}_{T}^{\dag}}^2=0,{\,}{\,}\{{\mathcal{Q}}_{T},\mathcal{W}_T\}=0,{\,}{\,}\mathcal{W}_T^2=I,{\,}{\,}[\mathcal{W}_T,H_{T}]=0
.\end{equation}
In this case, the Witten parity operator is equal to:
\begin{equation}\label{wparityopera}
\mathcal{W}_T= \left ( \begin{array}{cccc}
  1 & 0 & 0 & 0 \\
  0 & -1 & 0 & 0 \\
0 & 0 & 1 & 0 \\
0 & 0 & 0 & -1  \\
\end{array} \right)
.\end{equation}
Apart from this representation, there exist equivalent higher dimensional representations for the combined $N=2$, $d=1$ algebra, which can be obtained from the above algebra, by making the following sets of replacements:
\begin{equation}\label{setof transformations}
\mathrm{Set}{\,}{\,}{\,}A:{\,}
\begin{array}{c}
 {\mathcal{D}}_{RN}\rightarrow {\mathcal{D}}_{RN}^{\dag} \\
  {\mathcal{D}}_{RN'}^{\dag}\rightarrow {\mathcal{D}}_{RN'} \\
\end{array},{\,}{\,}{\,}\mathrm{Set}{\,}{\,}{\,}B:{\,}
\begin{array}{c}
 {\mathcal{D}}_{RN}\rightarrow {\mathcal{D}}_{RN'}^{\dag} \\
  {\mathcal{D}}_{RN'}^{\dag}\rightarrow {\mathcal{D}}_{RN} \\
\end{array},{\,}{\,}{\,}\mathrm{Set}{\,}{\,}{\,}C:{\,}
\begin{array}{c}
 {\mathcal{D}}_{RN}\rightarrow {\mathcal{D}}_{RN'} \\
  {\mathcal{D}}_{RN'}^{\dag}\rightarrow {\mathcal{D}}_{RN}^{\dag} \\
\end{array}
.\end{equation}
Moreover, a higher order reducible representation of the $N=2$ SUSY QM algebra supercharges, that is equivalent to the one materialized in relation (\ref{connectirtyrtons}), is given by: 
\begin{equation}\label{connectirtyfhfghrtons}
{\mathcal{Q}}_{T}= \left ( \begin{array}{cccc}
  0 & 0 & 0 & 0 \\
  0 & 0 & 0 & 0 \\
{\mathcal{D}}_{RN} & 0 & 0 & 0 \\
0 & {\mathcal{D}}_{RN'}^{\dag} & 0 & 0  \\
\end{array} \right),{\,}{\,}{\,}{\,}{\mathcal{Q}}_{T}^{\dag}= \left ( \begin{array}{cccc}
  0 & 0 & {\mathcal{D}}_{RN}^{\dag} & 0 \\
   & 0 & 0 & {\mathcal{D}}_{RN'} \\
0 & 0 & 0 & 0 \\
0 & 0 & 0 & 0  \\
\end{array} \right)
.\end{equation}
In addition, the Hamiltonian of the corresponding quantum system, is,
\begin{equation}\label{connectihgghdhtons1dtr}
H_T= \left ( \begin{array}{cccc}
  {\mathcal{D}}_{RN}^{\dag}{\mathcal{D}}_{RN} & 0 & 0 & 0 \\
  0 & {\mathcal{D}}_{RN'}^{\dag}{\mathcal{D}}_{RN'} & 0 & 0 \\
0 & 0 & {\mathcal{D}}_{RN'}{\mathcal{D}}_{RN'}^{\dag} & 0 \\
0 & 0 & 0 & {\mathcal{D}}_{RN}{\mathcal{D}}_{RN}^{\dag}  \\
\end{array} \right)
.\end{equation}
There exist other equivalent higher order representations, which we omit for brevity. In conclusion, there exists no extended $N\geq 3$ supersymmetric quantum algebra underlying the fermionic systems we have studied in this paper, only higher order reducible representations of the $N=2$ SUSY QM algebra, and this happens only in the case of the fermionic system in the Reissner-Nordstr\"{o}m-anti-de Sitter spacetime.

\subsection{A Global R-symmetry of the Hilbert Space of Quantum States}

Due to the $N=2$ SUSY QM algebra, the Hilbert space of quantum states possesses an inherent global R-symmetry, which is actually a $U(1)$. This is common to all systems we described in the previous sections, so we do not specify to a particular choice of supercharges. Suppose that the SUSY QM algebra is described by the supercharges ${\mathcal{Q}}_{G}$ and ${\mathcal{Q}}^{\dag}_{G}$. The $N=2$ superalgebra is
invariant under the transformations:
\begin{align}\label{transformationu1}
& {\mathcal{Q}}_{G}^{'}=e^{-ia}{\mathcal{Q}}_{G}, {\,}{\,}{\,}{\,}{\,}{\,}{\,}{\,}
{\,}{\,}{\mathcal{Q}}^{'\dag}_{G}=e^{ia}{\mathcal{Q}}^{\dag}_{G}
.\end{align}
Thus the quantum system is invariant under an $R$-symmetry of the
form of a global-$U(1)$. Obviously, the Hamiltonian of the SUSY QM algebra is invariant under the $U(1)$-transformation, that is $H_M'=H_M$. This R-symmetry has a direct impact on the transformation properties of the Hilbert states of the system. Recall that the total Hilbert space $\mathcal{H}$, is $Z_2$-graded. Let $\psi^{+}_{M}$ and
$\psi^{-}_{M}$ denote the Hilbert states corresponding to the
spaces $\mathcal{H}^{+}_{M}$ and $\mathcal{H}^{-}_{M}$. The even and odd states, are transformed under the
$U(1)$ transformation, as follows:
\begin{equation}
\psi^{'+}_{M}=e^{-i\beta_{+}}\psi^{+}_{M},
{\,}{\,}{\,}{\,}{\,}{\,}{\,}{\,}
{\,}{\,}\psi^{'-}_{M}=e^{-i\beta_{-}}\psi^{-}_{M}
.\end{equation}
The parameters $\beta_{+}$ and $\beta_{-}$ are global
parameters defined in such a way that $a$ is equal to $a=\beta_{+}-\beta_{-}$. Therefore, the resulting quantum system possesses a continuous R-symmetry.

\noindent Let us comment here that the breaking of this R-symmetry into a discrete one, can be a very interesting situation. An intriguing question is, how to break this continuous symmetry explicitly, because an explicit breaking could make us assume that a supercharge acquires a constant vacuum expectation value (assuming it's vacuum expectation value is it's trace over all vacuum eigenstates). Such a thing does not spoil the trace-class properties of the operators we studied and it's physically appealing since discrete symmetries are inherent to quantum system with fermionic condensates. Although interesting, such a task exceeds the purposes of this article.

\subsection{A Global two term Spin Complex Structure}

In this subsection, we shall present a spin complex structure, inherent to every fermionic system in the gravitational backgrounds we described in the previous sections. As we have seen, the Witten parity $\mathcal{W}$ provides a $Z_2$ grading to the Hilbert space of the SUSY QM mechanics algebra. Hence, the Hilbert space of each gravitational fermionic system is split in the following way $\mathcal{H}(M)=\mathcal{H}^+(M)\oplus \mathcal{H}^-(M)$, with $M$ denoting the corresponding curved spacetime. Each of the $\mathcal{H}^{\pm}(M)$ contains the even and odd Witten parity states. This grading can be used to define a spin complex, with the aid of the superconnection, which as we will demonstrate later on in this section, is the supercharge. In order to see this, let us note an important implication of the SUSY QM algebra on the vectors that belong to the graded Hilbert spaces $\mathcal{H}^{\pm}(M)$. Recall the vectors $|\psi^{+}\rangle$ $\mathcal{H}^+(M)$ and $|\psi^{-}\rangle$ $\in$ $\mathcal{H}^-(M)$, which are equal to, 
\begin{align}\label{phi5gdfghfdh}
&|\psi^{+}\rangle =\left(%
\begin{array}{c}
  |\phi ^{+} \rangle \\
  0 \\
\end{array}%
\right),{\,}{\,}{\,}\in{\,}{\,}\mathcal{H}^+(M)
\\ \notag &
|\psi^{-}\rangle =\left(%
\begin{array}{c}
  0 \\
  |\phi ^{-} \rangle \\
\end{array}%
\right),{\,}{\,}{\,}\in{\,}{\,}\mathcal{H}^-(M),
\end{align}
with $|\phi^{\pm}\rangle$, the vectors corresponding to the matrices $\mathcal{D}_i$, defined in the previous sections. As can be easily checked, the supercharges ${\mathcal{Q}}$, ${\mathcal{Q}}^{\dag}$ of the SUSY QM algebra,
\begin{equation}\label{wit2dyjtyjtyr1}
{\mathcal{Q}}=\bigg{(}\begin{array}{ccc}
  0 & {\mathcal{D}}_{i} \\
  0 & 0  \\
\end{array}\bigg{)}, {\,}{\,}{\,}{\,}{\,}{\mathcal{Q}}^{\dag}=\bigg{(}\begin{array}{ccc}
  0 & 0 \\
  {\mathcal{D}}_{i}^{\dag} & 0  \\
\end{array}\bigg{)},
\end{equation}
act in the following way on the vectors $|\psi^{\pm}\rangle$:
\begin{align}\label{wittyuhjhjhhtyi2}
&\bigg{(}\begin{array}{ccc}
  0 & \mathcal{D}_i \\
  0 & 0  \\
\end{array}\bigg{)}\left(%
\begin{array}{c}
   0\\
  |\phi ^{-} \rangle \\
\end{array}\right)=\left(%
\begin{array}{c}
  |\phi ^{-} \rangle \\
  0 \\
\end{array}\right),{\,}{\,}{\,}\in{\,}{\,}\mathcal{H}^+(M)
\\ \notag & \bigg{(}\begin{array}{ccc}
  0 & 0 \\
  \mathcal{D}_i^{\dag} & 0  \\
\end{array}\bigg{)}\left(%
\begin{array}{c}
  |\phi ^{+} \rangle \\
  0 \\
\end{array}\right)=\left(%
\begin{array}{c}
   0\\
  |\phi ^{+} \rangle \\
\end{array}\right),{\,}{\,}{\,}\in{\,}{\,}\mathcal{H}^-(M)
.\end{align}
Therefore, the supercharges imply the following two maps:
\begin{align}\label{mapsscharge}
&{\mathcal{Q}}:\mathcal{H}^-(M)\rightarrow \mathcal{H}^+(M)
\\ \notag & {\mathcal{Q}}^{\dag}:\mathcal{H}^+(M)\rightarrow \mathcal{H}^-(M)
.\end{align}
These two maps in turn, constitute a two term spin complex which has the following form:
$$\harrowlength=40pt \varrowlength=.618\harrowlength
\sarrowlength=\harrowlength
\mathcal{H}^+(M)\commdiag{\mapright^{{\mathcal{Q}}^{\dag}} \cr \mapleft_{{\mathcal{Q}}}}\mathcal{H}^-(M)$$
The index of this two term spin complex, is equal to the index of the operator $\mathcal{D}_i$, that is, $\mathrm{ind}_t\mathcal{D}_i$.

\subsection{The Krein Spectral Shift}

\noindent From the Witten index of the fermionic gravitational systems we studied in this paper, we can directly compute other topological quantities related to the corresponding spectral problems. Particularly, we shall be interested in the Krein
spectral shift $\xi(\lambda)$ \cite{thaller}. It worths recalling the definition of this function. Following \cite{thaller}, given two trace class operators $T_1$ and $T_2$, the function $\xi:\Re\rightarrow \Re$ is defined in such a way such that:
\begin{align}\label{xifunction}
&\mathrm{tr}(T_1-T_2)=\int_{-\infty}^{\infty}\xi(\lambda )\mathrm{d}\lambda
\\ \notag & \mathrm{tr}(f(T_1)-f(T_2))=\int_{-\infty}^{\infty}\xi(\lambda )f'(\lambda )\mathrm{d}\lambda.
\end{align}
The regularized index of an operator $\mathcal{Q}$, is written in terms of the Krein spectral shift as follows:
\begin{equation}\label{specshiftrela}
\mathrm{ind}_t\mathcal{Q}=\int_0^{\infty}\frac{t\xi (\lambda )}{(\lambda-t)^2}\mathrm{d}\lambda.
\end{equation}
Recall from relation (\ref{heatkerw}), that the Witten index is the limit of the regularized index for $t \rightarrow \infty$. It is proven that the Witten index is equal to $\xi (\infty)$ \cite{thaller}, that is:
\begin{equation}\label{}
\Delta_t=\xi (\infty ).
\end{equation}
Since the Witten index for the fermionic gravitational systems we studied is zero, we easily obtain that the corresponding Krein spectral shift is zero, that is
\begin{equation}\label{gfdghdfhfh}
\xi (\infty )=0.
\end{equation}

\subsection{Compact Radial Perturbations of Maximally SUSY QM and the Witten Index}

In the previous sections we found that an $N=2$ SUSY QM underlies fermionic systems when these are considered in curved gravitational backgrounds. As we will demonstrate, this SUSY QM algebra underlies any massless Dirac fermionic system when this is considered in a $D$-dimensional maximally symmetric spacetime. In addition, what we are mainly interested here, is to see what is the impact of compact radial perturbations to the Witten index of the SUSY QM algebra. 

\noindent We start first with the $N=2$ SUSY QM issue. A $D$-dimensional maximally symmetric spacetime ($D\geq 4$), has the metric of the form \cite{ortega}:
\begin{equation}\label{mmaximmetricsymm}
\mathrm{d}s^2=F(r)^2\mathrm{d}t^2-G(r)^2\mathrm{d}r^2-H(r)^2\mathrm{d}\Sigma^2_{D-2}
.\end{equation}
The corresponding Dirac equations for this background are \cite{ortega}:
\begin{align}\label{maxsymmbacjkoroyfnd}
&\Big{(}\partial_t-\frac{F(r)}{G(r)}\partial_r\Big{)}\psi_2=i\kappa \frac{F(r)}{H(r)}\psi_1
\\ \notag & \Big{(}\partial_t+\frac{F(r)}{G(r)}\partial_r\Big{)}\psi_1=-i\kappa \frac{F(r)}{H(r)}\psi_2
.\end{align}
Following the lines of research of the previous sections, an unbroken $N=2$ SUSY QM algebra underlies this fermionic system. The supercharges of this SUSY algebra are:
\begin{equation}\label{wit2dfdgdfgr1}
{\mathcal{Q}}_{M}=\bigg{(}\begin{array}{ccc}
  0 & {\mathcal{D}}_{M} \\
  0 & 0  \\
\end{array}\bigg{)}, {\,}{\,}{\,}{\,}{\,}{\mathcal{Q}}^{\dag}_{M}=\bigg{(}\begin{array}{ccc}
  0 & 0 \\
  {\mathcal{D}}_{M}^{\dag} & 0  \\
\end{array}\bigg{)}
,\end{equation}
and the Hamiltonian is
\begin{equation}\label{wit4354drgdgdfgrt1231}
H_{M}=\bigg{(}\begin{array}{ccc}
  {\mathcal{D}}_{M}{\mathcal{D}}_{M}^{\dag} & 0 \\
  0 & {\mathcal{D}}_{M}^{\dag}{\mathcal{D}}_{M}  \\
\end{array}\bigg{)}
,\end{equation}
with ${\mathcal{D}}_{M}$, the operator:
\begin{equation}\label{susyqmrgdfgfhfhn567m}
{\mathcal{D}}_{M}=\left(%
\begin{array}{cc}
 \partial_t-\frac{F(r)}{G(r)}\partial_r & i\kappa \frac{F(r)}{H(r)}
 \\  -i\kappa \frac{F(r)}{H(r)} & \partial_t+\frac{F(r)}{G(r)}\partial_r\\
\end{array}%
\right).
.\end{equation}
In the above, $\kappa$ stands for the eigenvalues of the Dirac operator in the $D-2$-dimensional manifold $\mathrm{d}\Sigma ^2_{D-2}$. It can be easily shown that the Witten index of the SUSY QM algebra is $\Delta  =0$, with $\mathrm{ker}{\mathcal{D}}_{M}=\mathrm{ker}{\mathcal{D}}_{M}^{\dag}\neq 0$. Hence supersymmetry is unbroken.

\noindent A perturbation of this fermionic system is materialized by adding a function to the term $W=i\kappa \frac{F(r)}{H(r)}$, of the form:
\begin{equation}\label{cp}
W'(r)=W(r)+S(r)
.\end{equation}
We employ a simplified approach to perturbations, since we are not interested for the perturbations, but for the index of the operator ${\mathcal{D}}_{M}$. A much more detailed analysis on perturbations can be found in \cite{ortega}. The function $S(r)$ is considered to be  fast convergent, so that the following operator is compact:
\begin{equation}\label{susyqmshadrnmassive}
C_M=\left(%
\begin{array}{cc}
 0 & S(r)
 \\ S(r) & 0\\
\end{array}%
\right)
.\end{equation}
Hence, since we can write ${\mathcal{D}}_{M}'={\mathcal{D}}_{M}+C_M$, and the operator (\ref{susyqmshadrnmassive}) is compact and odd, according to the theorem of section 2, the index of the operator ${\mathcal{D}}_{M}$, is invariant:
\begin{equation}\label{indperturbatrn}
\mathrm{ind}_{t}({\mathcal{D}}_{M}+C_M)=\mathrm{ind}_{t}{\mathcal{D}}_{M}
.\end{equation}
Therefore, the Witten index does not change under compact odd perturbations of the fermionic system. This means that $N=2$, $d=1$ SUSY is unbroken in the case of compact perturbations.

\subsection{Some Local Geometric Implications of the SUSY QM Algebra on the Spacetime Fibre Bundle Structure}

In this section, we shall present in brief the local geometric implications of the SUSY QM algebra on the fibre bundle structure of the spacetime $M$. This local geometric structure is a common attribute of all the spacetimes we studied in the previous sections. Particularly, we shall demonstrate that locally, the spacetime manifold $M$ (and locally means at an infinitesimally small open neighborhood of a point $x$ $\in$ $M$), due to the $N=2$ SUSY QM algebra, is rendered a supermanifold, with the supercharge of the SUSY QM algebra being the superconnection on this supermanifold, and the square of the supercharge being the corresponding curvature. For some important references on the mathematical issues we will use see \cite{graded1,graded2,graded3,graded4,graded5,graded6,graded7,graded8,graded9,graded10,graded11,graded12,graded13,graded14}.

\noindent The charged fermions that are defined on the spacetime $M$, are sections of the total $U(1)$-twisted fibre bundle $P\times S \otimes U(1)$, where $S$ is the representation of the Spin group $Spin(4)$, which in four dimensions is reducible, and $P$, the double cover of the principal $SO(4)$ bundle on the tangent manifold $TM$. Recall that a $Z_2$ grading on a vector space $E$, is a decomposition of the vector space $E=E_+\oplus E_{-}$. A decomposition of an algebra $A$, to even and odd elements, $A=A_+\oplus A_{-}$, such that: 
\begin{equation}\label{amodule}
A_+\cdot E_+\subset E_+,{\,}{\,}A_+\cdot E_-\subset
E_-,{\,}{\,}A_-\cdot E_+\subset E_-,{\,}{\,}A_-\cdot E_-\subset
E_+,
\end{equation}
is called a $Z_2$-grading of $A$, and the algebra $A$ is called a $Z_2$-graded algebra. Let the vector space $E=E_+\oplus E_{-}$, and $\mathcal{W}$ an involution $\in$ $\mathrm{End} (E)$, that is, it belongs to the endomorphisms of $E$. In addition, it is assumed that $\mathcal{W}=\pm1$ on the vectors of $E$, that is:
\begin{equation}\label{befoend}
\mathcal{W}(a+b)=a-b,{\,}{\,}{\,}\forall{\,}{\,}a{\,}\in{\,}E_+,{\,}{\,}\mathrm{and}{\,}{\,}\forall{\,}{\,}b{\,}\in{\,}E_-.
\end{equation} 
This involution $\mathcal{W}$, renders the algebra $\mathrm{End} (E)$ a $Z_2$ algebra. In the case at hand, for the vector space $\mathcal{H}$, the elements of $\mathrm{End} (\mathcal{H})$ are matrices of the form:
 \begin{align}\label{evenodd}
&\bigg{(}\begin{array}{ccc}
  0 & g_1 \\
  g_2 & 0  \\
\end{array}\bigg{)},{\,}{\,}{\,}\mathrm{odd}{\,}{\,}{\,}\mathrm{elements}
\\ \notag &
\bigg{(}\begin{array}{ccc}
 g_1 & 0 \\
  0 & g_2  \\
\end{array}\bigg{)},{\,}{\,}{\,}\mathrm{even}{\,}{\,}{\,}\mathrm{elements},
\end{align}
with $g_1,g_2$ complex numbers in general. The $N=2$ SUSY QM algebra,
$\mathcal{W},\mathcal{Q},\mathcal{Q}^{\dag}$ and particularly the
involution $\mathcal{W}$, generates the $Z_2$ graded vector space
$\mathcal{H}=\mathcal{H}^+\oplus \mathcal{H}^-$. The subspace
$\mathcal{H}^+$ contains $\mathcal{W}$-even vectors while $\mathcal{H}^-$, $\mathcal{W}$-odd vectors.
This grading is actually an additional algebraic structure $\mathcal{A}$ on $M$, with
$\mathcal{A}=\mathcal{A}^+\oplus \mathcal{A}^-$ an $Z_2$ graded
algebra. Particularly, $\mathcal{A}$ is a total rank $m$ ($m=2$ for
our case) sheaf of
$Z_2$-graded commutative $R$-algebras. Hence $M$
becomes a graded manifold $(M,\mathcal{A})$.

\noindent The sheaf $\mathcal{A}$ contains the endomorphism $\mathcal{W}$, $\mathcal{W}:\mathcal{H}\rightarrow
\mathcal{H}$, with $\mathcal{W}^2=I$. Thereby, $\mathrm{End}(\mathcal{H})\subseteq \mathcal{A}$. $\mathcal{A}$ is called a structure sheaf of the graded manifold
$(M,\mathcal{A})$, while $M$ is the body of
$(M,\mathcal{A})$. 

\noindent The structure sheaf $\mathcal{A}$ is isomorphic to the sheaf
$C^{\infty}(U)\otimes\wedge R^m$ of some exterior affine vector bundle
$\wedge \mathcal{H_E}^*=U\times \wedge R^m$, with $\mathcal{H_E}$
the affine vector bundle with fiber the vector space $\mathcal{H}$. The structure sheaf
$\mathcal{A}=C^{\infty}(U)\otimes\wedge \mathcal{H}$, is
isomorphic to the sheaf of sections of the exterior vector bundle
$\wedge \mathcal{H_E}^*=R\oplus
(\oplus^{m}_{k=1}\wedge^k)\mathcal{H_E}^*$. Without getting into much more detail to the sheaf structure of $(M,\mathcal{A})$, let us see the local geometric implications of the aforementioned sheaf structure. In the case at hand, the sections of the bundle $TM^*\otimes\mathcal{H}$ are actually those sections of the fermionic bundle $P\times S \otimes U(1)$, related to the SUSY QM algebra, which are the ones related to the radial part of the Dirac equation. A superconnection, denoted $\mathcal{S}$, is an one form with values in $\mathrm{End} (E)$ (or equivalently a section of  $TM^*\otimes \wedge \mathcal{H_E}^*\otimes
\mathcal{H_E}$). The curvature of the superconnection, which we denote $\mathcal{C}$, is a $\mathrm{End} (E)$-valued two form on $M$, such that $\mathcal{C}=\mathcal{S}^2$. Locally on $M$, this suggests that the superconnection is a section of $TM^*\otimes \mathrm{End} (E)^{\mathrm{odd}}$, which are simply the odd elements of $\mathrm{End} (E)$. 

\noindent Locally, the supercharge of the SUSY QM algebra is the superconnection, that is $\mathcal{S}={\mathcal{Q}}$, and therefore, the curvature of the supermanifold is locally $\mathcal{C}={\mathcal{Q}}^2$. 

\noindent In conclusion, the SUSY QM renders each manifold $M$ a graded manifold and locally a supermanifold, with superconnection ${\mathcal{Q}}$ and curvature ${\mathcal{Q}}^2$.

\noindent Finally, let us note that in addition to the local geometric implications we presented above, there are some global implications for the graded manifold $(M,A)$, but due to the complexity of their structure are, by far, out of the scopes of this article.

\subsection{Is There any Connection of SUSY QM and Spacetime Supersymmetry?}

With the spacetime $M$ being locally a supermanifold, it is unavoidable to ask whether there is some connection of the SUSY QM algebra with a global spacetime supersymmetry. The answer is no. When studying supersymmetric algebras in different dimensions we have to be cautious since the spacetime supersymmetric algebra (related to the super-Poincare algebra in four dimensions) is four dimensional while the SUSY QM algebra is one dimensional. Furthermore, spacetime supersymmetry in $d>1$ dimensions and
SUSY QM, that is $d=1$ supersymmetry, are different. There is however a connection, owing to the fact that extended (with $N = 4, 6...$) SUSY QM models describe
the dimensional reduction to one dimension of $N = 2$ and $N = 1$
Super-Yang Mills models \cite{ivanov1,ivanov2,ivanov3,ivanov4,ivanov5,ivanov6,ivanov7,ivanov8,ivanov9}. Still, the $N = 2$, $d=1$ SUSY QM supercharges
do not generate spacetime supersymmetry and therefore, SUSY QM does not relate fermions and bosons in terms of representations of the Poincare algebra in four dimensions. The SUSY QM supercharges provide a $Z_2$ grading on the Hilbert space of quantum states and also generate transformations between the Witten parity eigenstates. That is why the manifold $M$ is globally a graded manifold and not a supermanifold.

\section{Concluding Remarks}

In this paper we studied the relation of the quasinormal modes to SUSY QM. In particular, we showed that an $N=2$ SUSY QM algebra underlies Dirac fermionic systems when these are considered in various curved spacetimes backgrounds. We examined thoroughly the massless Dirac fermion Kerr black hole, the massless and massive fermion case in Kerr-Newman, Reissner-N\"{o}rdstrom, Schwarzschild, D-dimensional de Sitter, and Kerr-Newman-de Sitter spacetimes. In all the cases we found a hidden $N=2$, $d=1$ unbroken supersymmetry in the radial part of the Dirac fermion system, and the non-breaking of supersymmetry was very closely connected to the operators being trace-class and to the existence of quasinormal modes. Actually the zero modes of the fermionic system in each case, were in bijective correspondence to the quasinormal modes of each system.

\noindent Up to date, the important theoretical issue that addresses the nature of neutrino, that is whether it is Dirac or Majorana, has not be answered successfully yet. Hence, any information on the effect of massless (if the neutrino can be considered massless) fermions in nature is invaluable. We studied massless Dirac fermion in realistic curved four dimensional gravitational backgrounds and found an underlying one dimensional supersymmetry. This result provides us with valuable information, in view of the fact that the existence of quasinormal modes guarantees the unbroken supersymmetry.

\noindent In view of gauge/gravity dualities, it would be interesting to search for such supersymmetries in quantum systems with AdS backgrounds. In addition, such a study would be complete if both Majorana and Dirac fermion zero modes are studied in such backgrounds, in reference to supersymmetric structures. Moreover, this would require the Majorana fermion quasinormal mode spectrum, in black hole backgrounds, and since the nature of the neutrino is not known (that is whether it is Majorana or Dirac), up to date, it would be interesting by it self to study this issue.  

\noindent Finally, it worths investigating whether there is any connection between the topology of the black holes and the $N=2$ SUSY QM algebras we found. This connection to the topology is also advocated by the fact that, the Witten index (a topological property characterizing the various spin structures of the spacetime) of the algebra vanishes, but the kernels of the corresponding operators are non-empty.

\end{document}